\documentclass[11pt,twocolumn]{article}
\usepackage{epsfig}
\usepackage{latexsym}
\makeatletter
\def\section{\@startsection{section}{1}{\z@}{-3.5ex plus-1ex minus-.2ex}%
{2.3ex plus.2ex}{\large\bf}}
\makeatother

\begin{document}
\twocolumn[\title{Energy-level statistics and localization of 2d electrons in
\\ random magnetic fields}
\author{M. Batsch$^{a,b}$, L. Schweitzer$^{a,}$\footnotemark[1] , B. 
Kramer$^b$\\
{\itshape $^a$Physikalisch-Technische Bundesanstalt, Bundesallee 100,}\\
{\itshape D-38116 Braunschweig, Germany}\\
{\itshape $^b$I. Institut f\"ur Theoretische Physik, Universit\"at Hamburg,
Jungiusstra{\ss}e 9,}\\{\itshape D-20355 Hamburg, Germany}}
\date{}
\maketitle]
\footnotetext[1]{Corresponding author: Ludwig.Schweitzer@ptb.de}
\begin{abstract}
Using the method of energy-level statistics, the localization properties of 
electrons moving in two dimensions in the presence of a perpendicular random 
magnetic field and additional random disorder potentials are investigated. 
For this model, extended states have recently been proposed to exist in the 
middle of the band.   
In contrast, from our calculations of the large-$s$ behavior of the 
nearest neighbor level spacing
distribution $P(s)$ and from a finite size scaling analysis we find only 
localized states in the suggested energy and disorder range.
\end{abstract}

\section{Introduction}
Many investigations, both analytical and numerical, have been undertaken to 
elucidate the localization problem of non-interacting electrons in random 
magnetic fields (RMF) (see e.g., \cite{SN93,LC94,AMW94,SW95}).
The RMF model has been proposed for studying the quantum Hall effect with 
Landau level filling $\nu=1/2$ where the system of quasi-particles may be 
viewed as a Fermi liquid like metal. 
In a mean field approximation these composite fermions \cite{HLR93,KZ92,Jai89}
are subject to a fluctuating local magnetic field with zero mean.

However, up to now, no satisfactory agreement has been reached on the 
question, whether or not a metal-insulator transition exists in the 
thermodynamic limit. In recent numerical studies \cite{SW95,YB96a}, 
current carrying states have been 
reported for a RMF model including additional random disorder potentials. 
The authors of Ref.~\cite{YB96a} report that a metal-insulator transition has 
been detected in the band center at a critical disorder $W_c=3.0\pm 0.2$ 
by making use of the topological properties of the wave functions. 

The method of calculating the topological quantum number (Chern number) of the 
one-electron wave functions, which has proven to be a useful tool in the 
integer quantum Hall effect \cite{YB96}, considerably restricts the achievable
system size, $L/a<20$. 
On the other hand, using the method of energy-level statistics 
\cite{Sea93,SZ95,BSZK96,BS97}, 
system sizes of more than a factor of 10 larger are accessible. In this method,
the knowledge of the eigenfunctions is not necessary, because the information 
about localized, critical or extended states is gained from the spectral 
correlations.
Therefore, we report here on an independent check of whether the extended 
states found in Ref.\,\cite{YB96a} can really be expected in the thermodynamic 
limit or if their result must be attributed to finite size effects.

\section{Model and Method}
A one-band tight-binding model is used to describe non-interacting particles
moving in two dimensions in the presence of spatially fluctuating magnetic 
fields.
\begin{equation}
H=\sum_{<k\ne l>} V \exp(i\phi_{kl})|k\rangle\langle l|+\sum_k 
\varepsilon_k |k\rangle\langle l|
\end{equation}
The $|k\rangle$ are a complete set of lattice vectors connected with the sites
of a square lattice with linear size $L$ and lattice constant $a$, and $V$ is
taken as the unit of energy. The sum runs over nearest neighbors only. 
The phases $\phi_{kl}$ along the bonds are chosen 
such that the corresponding magnetic fluxes per plaquette, 
$\Phi_{k}=\phi_{k,k+a_x}+\phi_{k+a_x,k+a_x+a_y}+\phi_{k+a_x+a_y,k+a_y}+
\phi_{k+a_y,k}$, are distributed according to 
$-h_0\pi \le \Phi_{k} \le h_0\pi$ with probability density 
$p(\Phi_{k})=1/(2h_0)$. The strength of the flux was taken to be $h_0=0.5$.

The diagonal disorder potentials $\varepsilon_k$ are chosen at random from an 
interval $[-W, W]$ with uniform distribution, $p(\varepsilon_k)=1/(2W)$.
We considered system sizes $L/a$ in the range 20 to 200 and disorder values
$W/V$ between 1.5 and 3.5.
The energy eigenvalues within a small range ($|E|/V < 0.3$) in the middle of 
the tight-binding band are calculated using a Lanczos algorithm.
For each pair of parameters ($L, W$) about 250\,000 eigenvalues are 
accumulated taking different realizations of the disorder potentials.
We carefully checked that the results were independent of the width of the
energy interval chosen.

\begin{figure}[t]
\epsfxsize8.cm\epsfbox{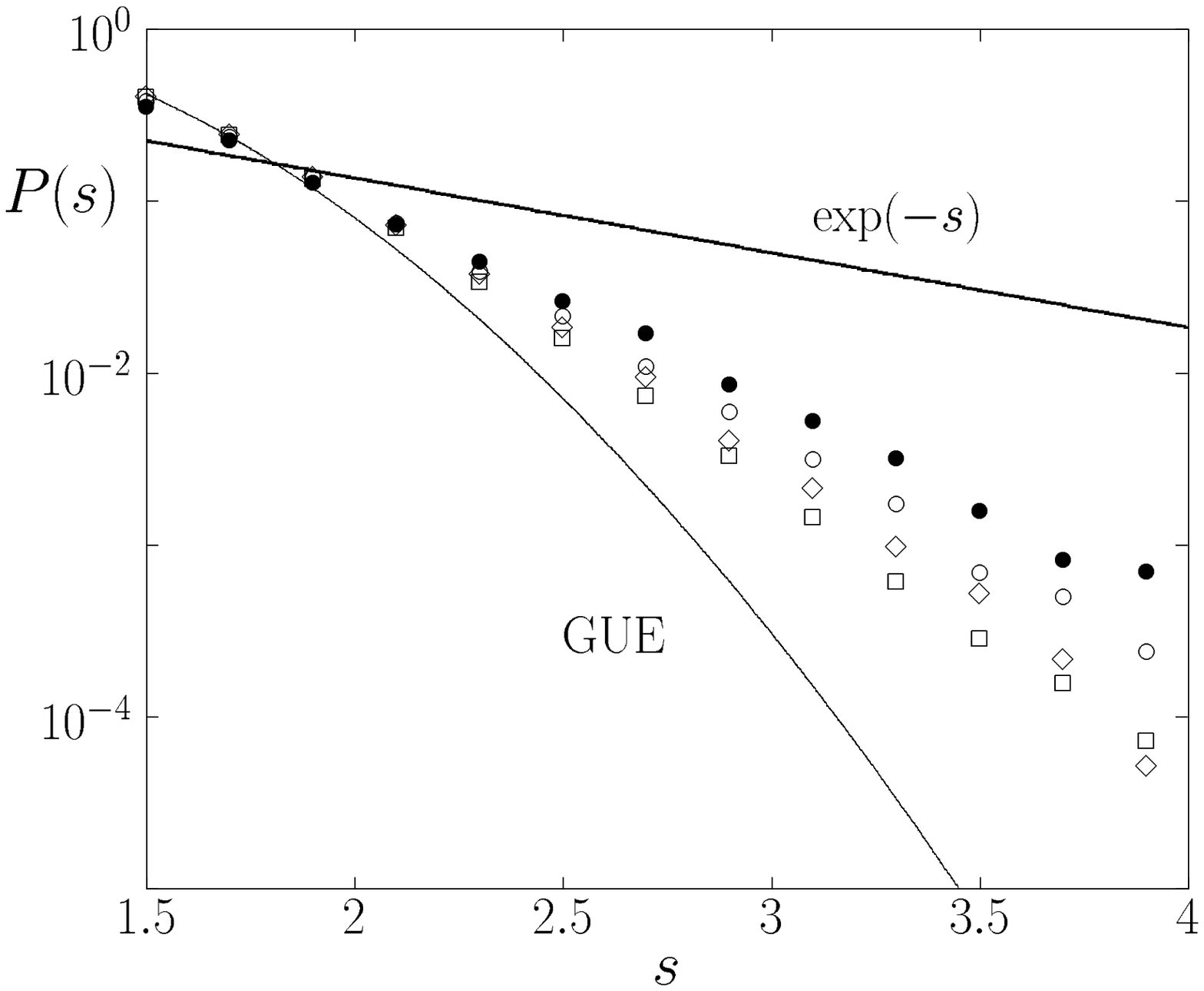}
\caption[]{\label{fig1}The large-s behavior of the nearest neighbor level 
spacing distribution $P(s)$ for different system sizes $L/a=20$ 
({\footnotesize$\Box$}),  
40 ($\diamond$), 100 ($\circ$), and 200 ($\bullet$). Disorder $W/V=2.75$ and 
strength of the random flux $h_0=0.5$.}
\end{figure}

From the unfolded energy eigenvalues the nearest neighbor level spacing 
distribution $P(s)$ is calculated, where $s=|E_{i+1}-E_{i}|/\Delta$ and 
$\Delta$ is the mean level spacing. For infinite systems it is known from 
random matrix theory (RMT) \cite{Meh91,Efe83} that in the metallic regime 
$P(s)$ is close to the Wigner surmise, 
$P(s)=A_\beta s^\beta \exp(-B_\beta s^2)$, where $\beta$
reflects the symmetry of the system: orthogonal ($\beta=1$), unitary 
($\beta=2$), and symplectic ($\beta=4$). In the insulating case the spacings
of the uncorrelated eigenvalues are well described by the Poisson probability 
density distribution $P(s)=\exp(-s)$.

In finite systems, when the correlation length exceeds the system size, 
one expects a hybrid \cite{Sea93} of the Wigner and the Poisson form which 
eventually tends to the limiting true metallic case, if $W<W_c$, or to 
a complete insulating behavior, if $W>W_c$, as the system size is increased, 
$L\to\infty$. 
For disorder strength $W$ close to the critical value $W_c$ various new 
scale independent distributions have been discovered recently in 
several systems \cite{Sea93,SZ95,BS97,KOSO96,ZK97} that exhibit a 
metal-insulator-transition or at least a critical point.
 
\begin{figure}[b,h]
\epsfxsize8.cm\epsfbox{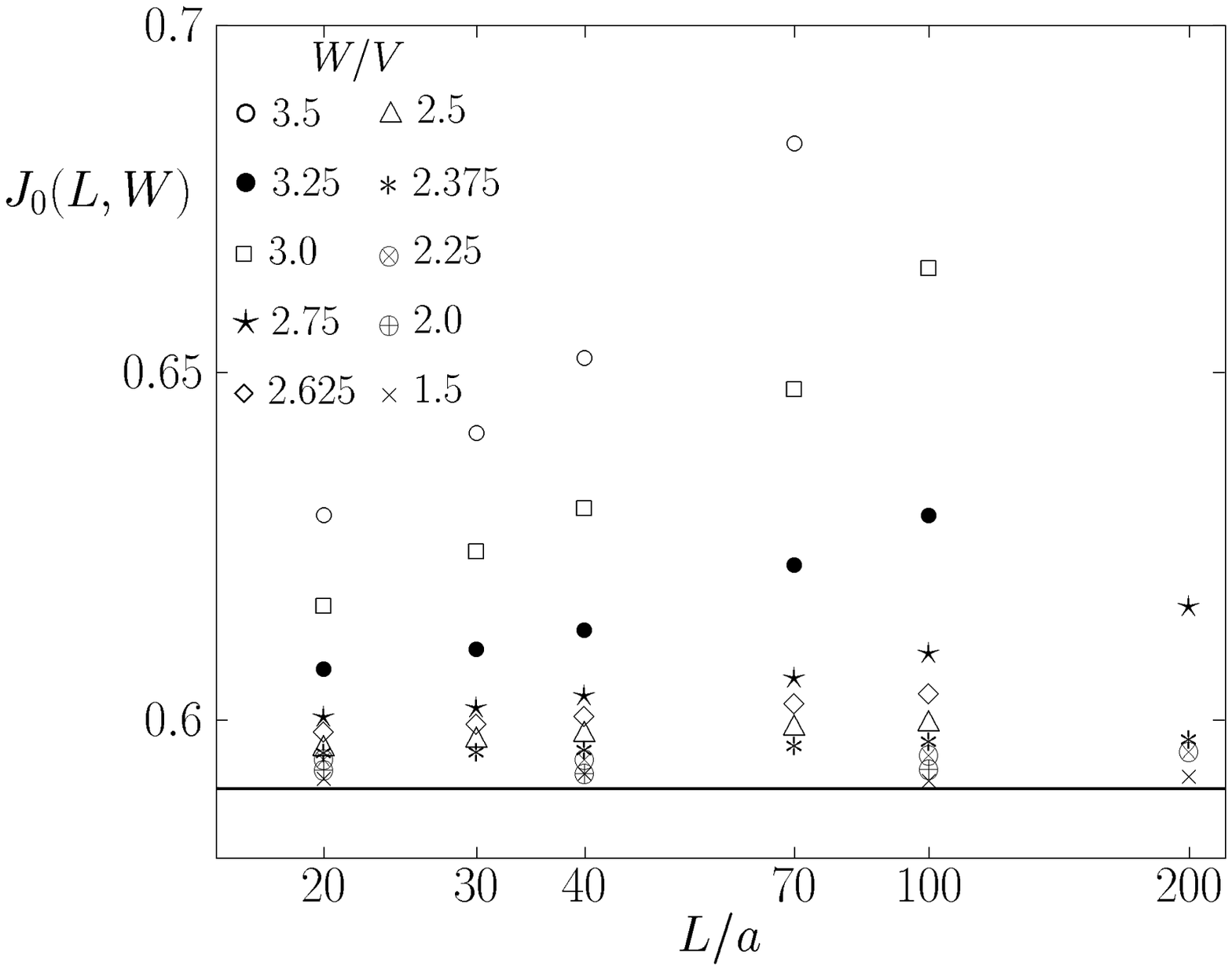}
\caption[]{\label{fig2}The calculated $J_0(L,W)$ for various system sizes $L$
and disorder strength $W$ showing a behavior expected for localized states. }
\end{figure}

\section{Results and Discussion}
In the 2d random flux model with additional diagonal disorder potentials, 
we find for small $s$, $P(s)\sim s^2$, as expected for broken time reversal 
symmetry (unitary). Fig.~\ref{fig1} shows only the large-s behavior of $P(s)$ 
for a disorder strength $W=2.75\,V$, which is smaller than the proposed 
critical disorder $W_c=3\,V$, and for system sizes $L/a=20$, 40, 100, and 200. 
The RMT result for the Gaussian unitary
ensemble (GUE) which holds in the metallic case and the Poisson result 
applicable for the uncorrelated eigenvalues of the localized states are also 
plotted. Increasing the system size shifts the curves away from the hybrid 
$P(s)$ towards the $\exp(-s)$ decay of the localized
states. Hence, even for $W$ smaller than the suggested critical 
$W_c/V=3$, all the states still tend to localization in the limit $L\to 
\infty$.

\begin{figure}[t]
\epsfxsize8.cm\epsfbox{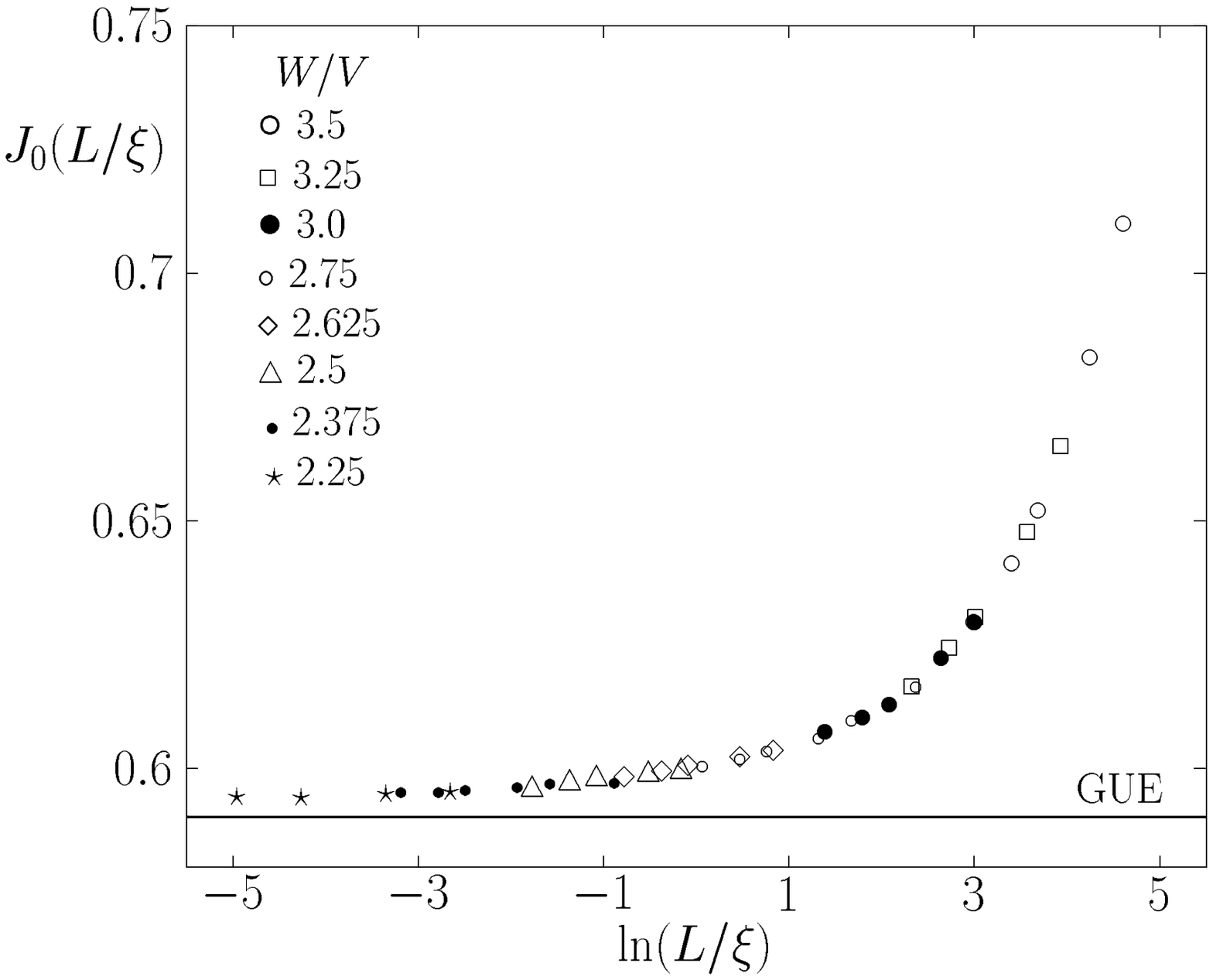}
\caption[]{\label{fig3}The result of the finite size scaling analysis is a
one-branch scaling function, $J_0(L/\xi)$. All states are localized for
disorder potential strength $2.25 \le W/V \le 3.5$.}
\end{figure}

To be more quantitative, we have calculated the number 
$J_0(L,W)=\int_0^\infty Q(0,s)\,ds$ where $Q(n,s)$ is the probability that a 
given energy interval $s$ contains exactly $n$ eigenvalues, and 
$P(s)=d^2 Q(0,s)/ds^2$. In the metallic regime, a value $J^x_0=0.590$ is known 
for infinite systems from RMT while for localized states $J^l_0=1$. 
In Fig.~\ref{fig2} $J_0(L,W)$ is shown
for different system sizes and disorder strength. The values clearly increase
with increasing $L/a$, at least for $W/V>2.25$. This indicates 
localized states at the band center for this disorder range. A weak decrease
of $J_0(L,W)$ is still found for smaller disorder strength, but no size 
dependence could be observed anymore, because in that case the localization 
length considerably exceeds our largest system size $L/a=200$. 

It is possible to use $J_0(L,W)$ also as a scaling variable, a procedure which
was exploited recently \cite{SZ97} to determine the critical exponent of the 
correlation function at the metal-insulator-transition (MIT) in a 2d 
symplectic system. For the RMF model we have tested the validity of the 
single parameter hypothesis $J_0(L,W)=f(L/\xi(W))$ of a finite size scaling 
analysis which was performed for the data showing a size dependence 
$(W/V\ge 2.25)$. The result is plotted in Fig.~\ref{fig3} where only a single 
branch of the scaling function can be seen which excludes the existence of a 
MIT for the disorder strengths investigated. 

\begin{figure}[bh]
\epsfxsize8.cm\epsfbox{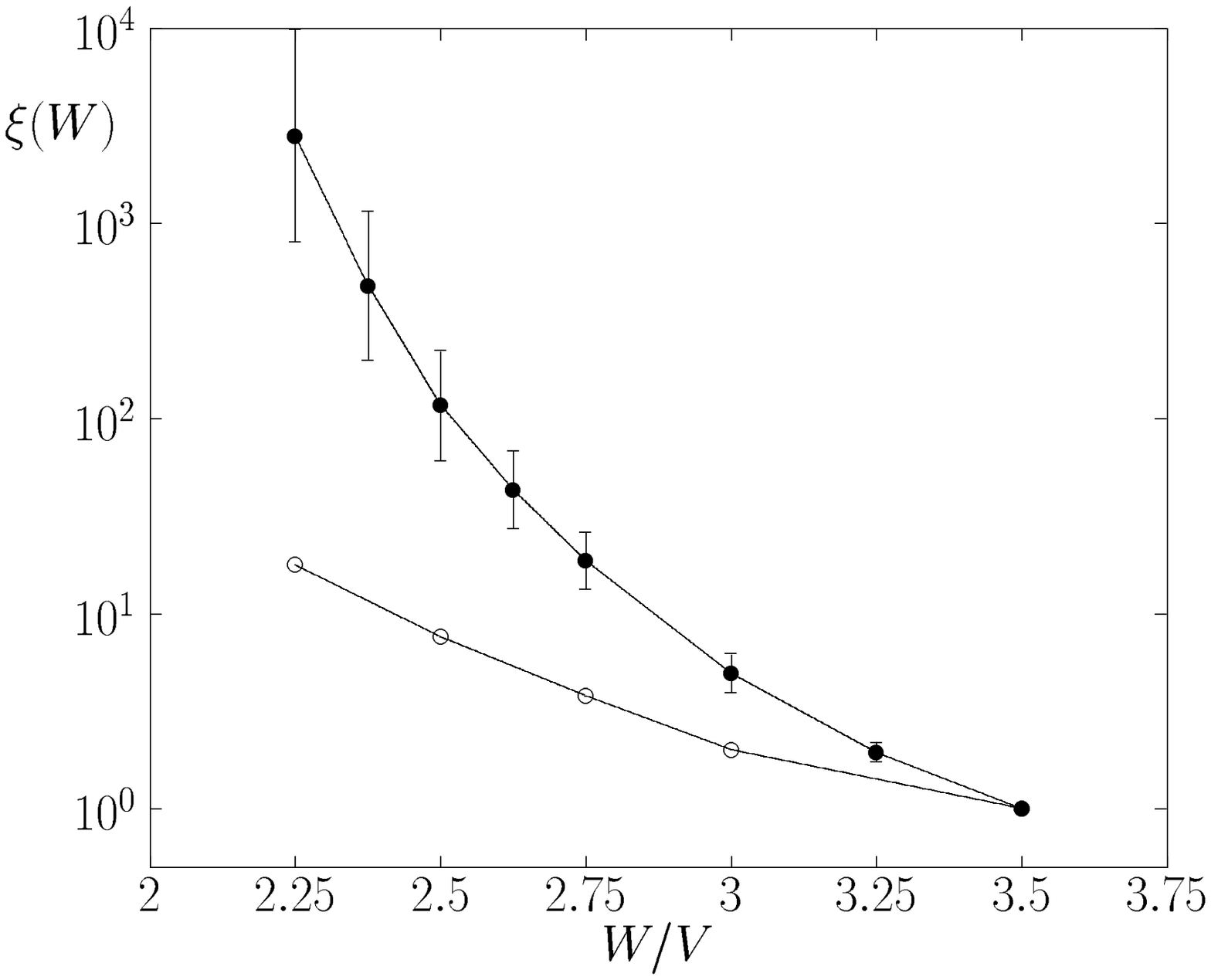}
\caption[]{\label{fig4}Disorder dependence of the correlation function 
$\xi(W)$ of the RFM ($\bullet$) and of the 2d Anderson model ($\circ$) both 
normalized such that $\xi(W/V=3.5)=1$.}
\end{figure}

The disorder dependence of the scaling parameter $\xi(W)$ is
shown in Fig.~\ref{fig4} in comparison with the result for a
conventional 2d Anderson model without random magnetic fields. 
Since the prefactor $\xi_0$ was not determined in our calculation,
both curves were set to 1 at $W/V=3.5$. The faster increase of $\xi(W)$ in the
RFM model indicates a weakening of the localization due to the random magnetic 
fields. The increase of the localization length is, however, much larger than 
the factor of 2 expected from RMT \cite{SMMP91} for the breaking of the time 
reversal symmetry.  

In conclusion, the localization properties of a two-dimensional electronic 
model containing random magnetic fields and additional disorder potentials 
have been investigated numerically. 
Taking the same parameters as in Ref.\,\cite{YB96a} for the strength of the 
random flux, $h_0=0.5$, and the disorder potential $W/V=3$, 
but system sizes up to $L/a=200$, our data definitely show complete 
localization of the electronic states. Even for smaller disorder no sign of a 
delocalized phase could be detected.  

\section*{Acknowledgments}
This work was supported in part by the ISI Foundation 
(ESPRIT 8050 Small Structures). L.S. would like to thank the organizers for 
the stimulating workshop at Villa Gualino, Torino.


\end{document}